# Radially correlated partially coherent beams with a deterministic vortex structure

Xinru Lian[1,2,*], Yaning Zhou[1,2,*], Xin Liu[1,2,†], Zhao Zhang[1,2], Qian Chen[1,2], Keyu Zhou[1,2], Weining Wang[1,2], Chunhao Liang[1,2,†], Yangjian Cai[1,2,†], Jingsong Liu[1,2,†]

[1]Shandong Provincial Key Laboratory of Light Field Manipulation Physics and Applications & School of Physics and Optoelectronics, Shandong Normal University, Jinan 250358, China

[2]Collaborative Innovation Center of Light Manipulations and Applications, Shandong Normal University, Jinan 250358, China

*These authors contributed equally to this work.

[†]Corresponding Authors: *xinliu@sdnu.edu.cn*; *chunhaoliang@sdnu.edu.cn*; *yangjiancai@sdnu.edu.cn*; *liujs@sdnu.edu.cn*;

**Abstract**

Partially coherent beams have attracted considerable attention due to their intrinsic resilience against complex environmental perturbations. However, the intrinsic wavefront fluctuations make it fundamentally challenging to preserve well-defined orbital angular momentum during propagation. In this work, we propose and experimentally demonstrate a class of radially correlated, partially coherent beams that carry deterministic vortex structures, generated via optical conformal mapping from Cartesian to log-polar coordinates. The resulting beams exhibit a ring-shaped coherence distribution, characterized by low coherence in the radial direction and high coherence in the azimuthal direction. This unique feature of such a beam supports a well-defined deterministic vortex phase, thereby enabling the beam to preserve its ring-shaped coherence distribution during propagation through a focusing system. Our results provide new insights into the design of new partially coherent beams and may facilitate the development of applications in optical encoding, free-space information transmission, and ultrafast light-matter interactions.

**TOC:**

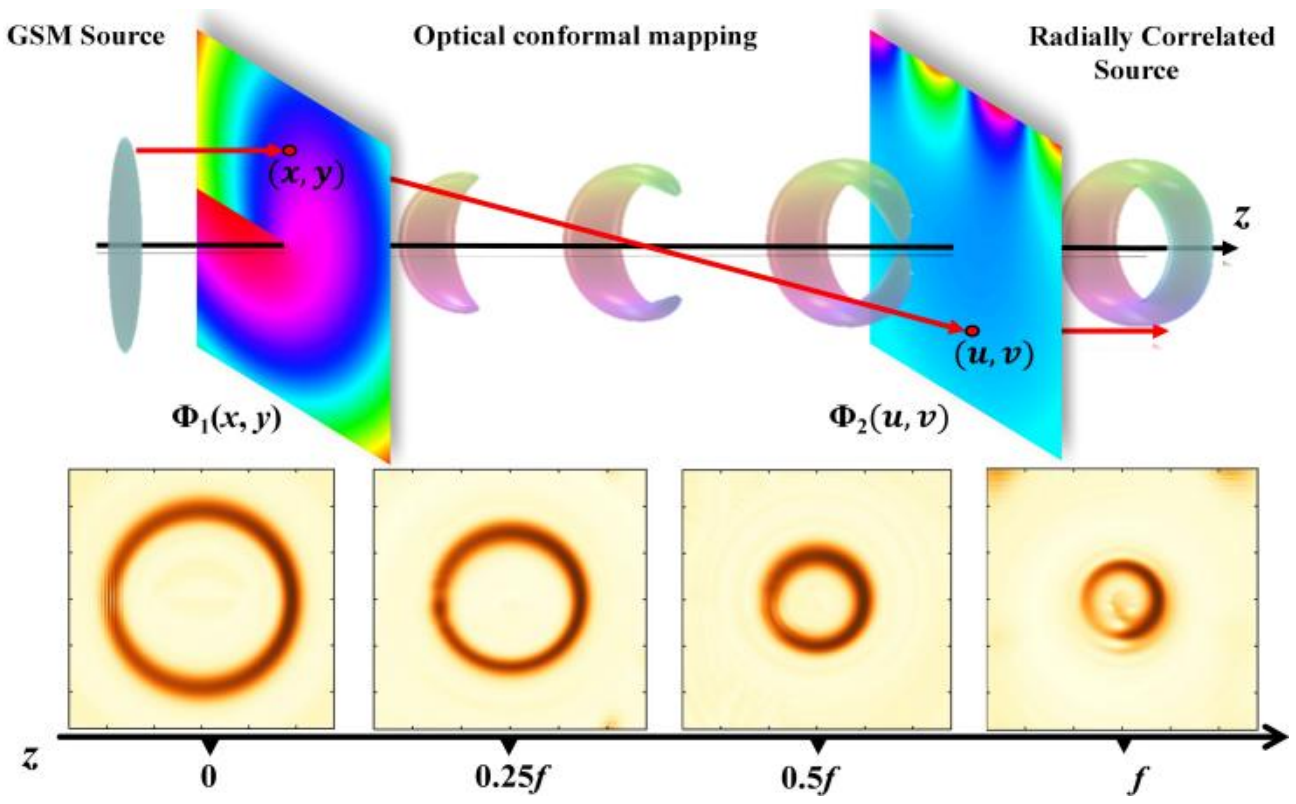

## Introduction

Partially coherent beams exhibit intrinsic robustness against random interference effects and environmental perturbations due to their statistical stability in amplitude and phase, offering distinct advantages in high-quality optical imaging [1,2], free-space optical communication [3,4], and high-performance photonic computing [5]. The coherence structure, as a second-order statistical quantity, measures the extent of statistical correlations of light fields at pairs of space-time points and determines the beam's propagation behavior [6,7]. Beyond conventional Gaussian Schell-model sources, precisely engineering coherence structures has enabled the generation of diverse structured optical fields with nonuniform correlation functions. These beams exhibit remarkable propagation characteristics, including self-focusing, self-shifting, self-healing, and diffraction-free propagation, leading to promising applications in optical trapping, optical manipulation, super-resolution imaging and free-space optical communication [8-18]. Notably, coherence structures exhibit superior robustness compared with first-order parameters such as amplitude and phase under turbulent and obstructed environments. Meanwhile, these statistical correlations require indirect characterization through appropriate measurement schemes. Leveraging this unique property, recent studies have demonstrated that coherence structures can be used as an additional information carrier, enabling image encryption and high-fidelity far-field information transmission under complex propagation conditions [19-24].

Vortex beams are a prominent class of structured light fields characterized by a helical phase wavefront, with each photon carrying an orbital angular momentum (OAM) of $l\hbar$, where $l$ is termed the topological charge [25]. Owing to unique OAM characteristics [26-30], vortex beams have attracted considerable attention for applications in optical manipulation, high-dimensional optical communications, and advanced imaging [31-34]. In recent years, partially coherent vortex beams have attracted increasing attention because they combine the enhanced robustness of light fields against environmental perturbations with the topological characteristics of optical vortex beams. Through the joint engineering of coherence distributions and phase structures, these beams can achieve enhanced resilience to complex environmental perturbations while offering increased information-carrying capacity, thereby opening new opportunities for advanced optical information transmission [35-37]. However, preserving deterministic OAM characteristics under low-coherence conditions remains a fundamental challenge. The underlying challenge arises from the intrinsic random phase fluctuations of partially coherent beams, which disrupt the preservation of their helical phase structure and lead to the progressive degradation of well-defined OAM during propagation. Recent studies have employed pseudo-mode superposition methods to generate partially coherent beams capable of preserving vortex characteristics in both focusing systems and free-space propagation [38,39]. Despite this progress, existing approaches remain strongly dependent on more specific modal configurations and have primarily focused on the evolution of intensity and OAM flux-density distributions. Consequently, developing more general strategies for constructing partially coherent beams with deterministic vortex structures, while clarifying how OAM governs the propagation dynamics of their coherence structures, is essential for advancing partially coherent vortex fields toward practical applications in optical information transmission.

In this work, we theoretically propose a family of radially correlated partially coherent beams and experimentally realize them through optical conformal mapping implemented with double-phase-screen modulation. The resulting beams exhibit a ring-shaped coherence distribution characterized by partial coherence in the radial direction and full coherence in the azimuthal direction. This distinctive coherence structure supports the stable propagation of a deterministic vortex phase structure and the preservation of well-defined longitudinal OAM, which in turn enables the beam to maintain its annular coherence profile during propagation.

## Theoretical analysis

Following the coherence theory, the coherence properties of light are second-order statistical quantities determined by the statistical correlations between stochastic optical fields evaluated at two spatial points, $\vec{\rho}_1 = (x_1, y_1)$ and $\vec{\rho}_2 = (x_2, y_2)$ [6]. In the space–frequency domain, a partially coherent beam is characterized by the cross-spectral density function associated with an ensemble of scalar random fields $\{E(\vec{\rho})\}$, which is defined as

$$\mathrm{W}(\vec{\rho}_1, \vec{\rho}_2) = \langle E(\vec{\rho}_1) E^*(\vec{\rho}_2) \rangle, \tag{1}$$

where $\langle\cdot\rangle$ denotes ensemble averaging, $\vec{\rho}_1$ and $\vec{\rho}_2$ represent arbitrary points in the transverse plane. For a physically realizable partially coherent optical field, the cross spectral density function satisfies the condition of non-negative definiteness and therefore can be represented in the form given by

$$\mathrm{W}(\vec{\rho}_1, \vec{\rho}_2) = \iint p(\vec{v}) H^*(\vec{\rho}_1, \vec{v}) H(\vec{\rho}_2, \vec{v}) d^2\vec{v}, \tag{2}$$

where $\boldsymbol{v} = (v_x, v_y)$ is an arbitrary point in the reverse space of $\vec{\rho} = (x, y)$, $p(\vec{v})$ is a non-negative real-valued function, and $H(\vec{\rho}, \vec{v})$ denotes an arbitrary kernel function. For a Gaussian Schell-model (GSM) source, these functions are read respectively defined as $p(\vec{v}) = \exp\left(-\delta_x{}^2 v_x{}^2 - \delta_y{}^2 v_y{}^2\right)$ and $H(\vec{\rho}, \vec{v}) = \exp\left(-x^2/w_x{}^2 - y^2/w_y{}^2\right) \exp\left(-2\pi i \vec{v} \cdot \vec{\rho}\right)$, where $\delta_x$ and $\delta_y$ characterize the transverse coherence widths, while $w_x$ and $w_y$ represents the beam waist along the transverse directions.

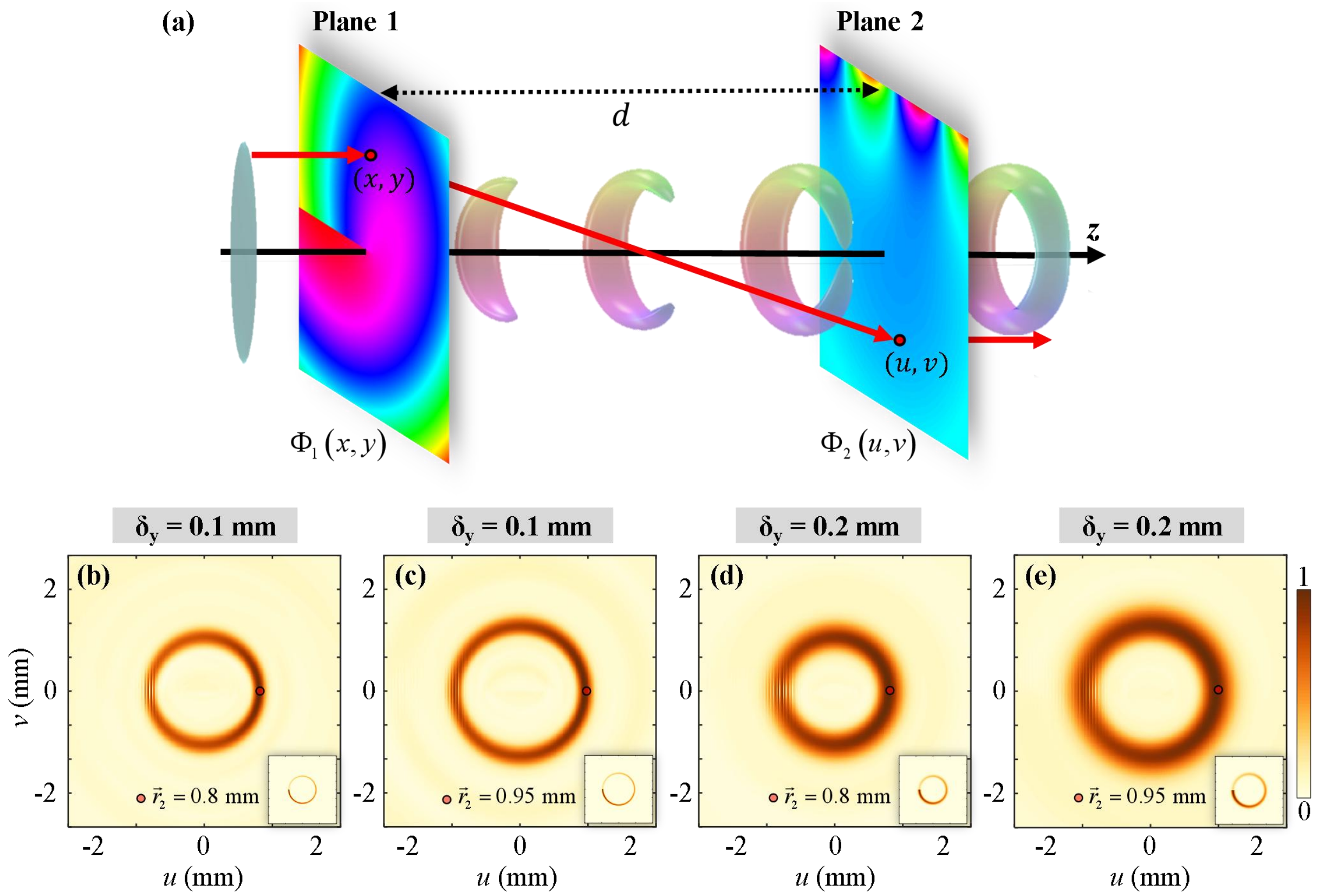

**Fig. 1.** (**a**) Schematic illustration of the conformal mapping scheme for generating radially correlated partially coherent beams with a coherence vortex structure. (**b**)-(**e**) Simulated and theoretical distributions of $|\mu(\vec{r}_1,\vec{r}_2)|^2$ for different coherent widths $\delta_y$ and reference points $\vec{r}_2$.

To construct radially correlated partially coherent beams, we firstly consider an initial GSM tube-like source that is fully coherent along the $x$-direction and partially coherent along the $y$-direction. The corresponding ratios of the beam waists and coherence widths are chosen as $w_x/w_y = 1{:}\,0.2$ mm and $\delta_x/\delta_y = 4{:}\,0.1$ mm, respectively. Subsequently, an optical coordinate transformation is introduced to map the Cartesian coordinate system into a log-polar coordinate system based on conformal mapping theory [40,41]. This transformation is realized using two cascaded phase screens, as illustrated in Fig. 1(**a**). Specifically, the beam first passes through the phase screen $\Phi_1(x,y)$, then propagates over a distance $z=d$, and finally traverses the second phase screen $\Phi_2(u,v)$. During the free-space propagation over the distance $d$ the phase modulation imposed by the first phase screen evolves into the field distribution required for the subsequent coordinate mapping. The second phase is used for alignment to compensate for the additional phase introduced by the transformation. The combined action of the two phase screens and intermediate propagation therefore accomplishes the optical coordinate transformation. In this conformal optical propagation system, the first phase mask implements the coordinate transformation from a tube geometry to a toroidal topology, and its phase profile can be analytically derived using the stationary phase approximation as

$$\Phi_1(x,y)=\frac{k}{d}\left[ab\exp\left(\frac{y}{a}\right)\sin\left(\frac{x}{a}\right)-\frac{\left(x^2+y^2\right)}{2}\right], \tag{3}$$

where $k$ is the wave number, $a$ and $b$ are scaling parameters that control the beam size and transverse position in the second plane. Following the same procedure, the phase profile of the second phase mask can be derived as

$$\Phi_2(u,v)=\frac{k}{d}\left[au\arctan\left(\frac{v}{u}\right)+\frac{av\ln\left(u^2+v^2\right)}{2}-av\left(\ln b+1\right)-\frac{\left(u^2+v^2\right)}{2}\right]. \tag{4}$$

After travelling through the two phase masks, the initial GSM tube-like source is transformed into a radially correlated partially coherent beam, whose CSD function can be approximately expressed as

$$W(\vec{r}_1,\vec{r}_2) = \exp\left\{-\frac{a^2(\theta_1^2+\theta_2^2)}{w_x^2} - \frac{a^2\left[(\ln r_1/b)^2 + (\ln r_2/b)^2\right]}{w_y^2}\right\}$$
$$\times \exp\left[-\pi^2 a^2 \frac{(\theta_1-\theta_2)^2}{\delta_x^2} - \pi^2 \frac{a^2}{b^2}\frac{(\ln r_1 - \ln r_2)^2}{\delta_y^2}\right], \quad (5)$$

where $\vec{r}_1 = (u_1, v_1)$ and $\vec{r}_2 = (u_2, v_2)$ denote two arbitrary position vectors in the second plane. $r = |\vec{r}| = \sqrt{u^2 + v^2}$ and $\theta = \arg(\vec{r}) = \arctan(v/u)$ are the radial and azimuthal coordinates, respectively, associated with the cylindrical coordinate system. The normalized form of the cross-spectral density function defines the complex degree of coherence, which is given by

$$\mu(\Delta\theta, \Delta \ln r) = \exp\left[-\pi^2 a^2 \frac{(\Delta\theta)^2}{\delta_x^2} - \pi^2 \frac{a^2}{b^2}\frac{(\Delta \ln r)^2}{\delta_y^2}\right]. \quad (6)$$

Figures 1(**b**)-1(**e**) show the simulated distributions of the squared modulus of the complex degree of coherence, $|\mu(\vec{r}_1,\vec{r}_2)|^2$, for the radially correlated partially coherent beam generated by the conformal mapping system illustrated in Fig. 1(**a**). The results are presented for different initial coherence widths $\delta_y$, and evaluated at various reference points $\vec{r}_2$. The insets show the corresponding theoretical predictions calculated from Eq. (6). It is evident that the $|\mu(\vec{r}_1,\vec{r}_2)|^2$ distribution exhibits a ring-shaped structure. The radius and width of the coherence ring are determined by the position of the reference point and the initial coherence width, respectively. The consistency between theory and simulation verifies the effectiveness of the proposed scheme for generating partially coherent beams with radially correlated coherence properties.

**Experimental analysis**

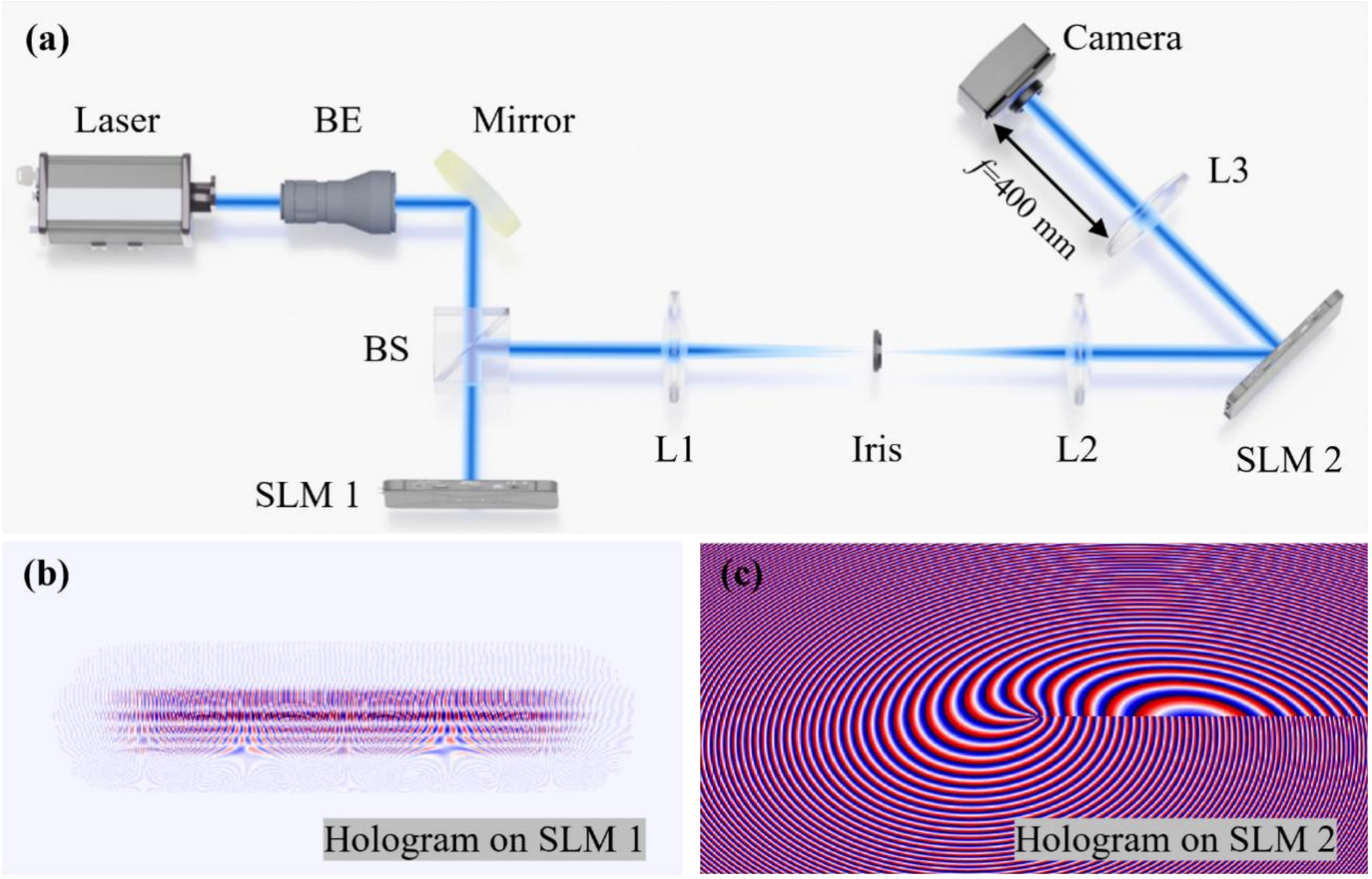

**Fig. 2.** (**a**) Experimental setup for generating radially correlated partially coherent beams with a deterministic vortex structure. BE, beam expander; BS, beam splitter; SLM 1-2, spatial light modulators; L1-3, lenses. (**b**) A representative hologram displayed on SLM 1, encoding the amplitude and phase of an instantaneous electric field, together with the first phase term required for conformal mapping. (**c**) A representative hologram displayed on SLM 2, encoding the second phase term for conformal mapping along with an additional helical phase.

In the experiment, the GSM source was first decomposed into a series of mutually uncorrelated random coherent modes using the random-mode representation method [42]. The resulting partially coherent beam was then synthesized as the incoherent superposition of these random modes, since the latter can be readily generated experimentally using digital holography techniques. Figure 2(**a**) shows the experimental setup for generating radially correlated partially coherent beams by means of the conformal mapping method. A laser beam emitted from a 532-nm laser is first expanded by a beam expander (BE) to obtain a collimated beam with a sufficiently large diameter. The expanded beam is then redirected by a mirror and directed onto a beam splitter (BS), which guides the beam to the first spatial light modulator (SLM 1). SLM1 displays a series of complex-encoded holograms corresponding to random modes $E(x,y)$, together with the first conformal mapping phase mask $\Phi_1(x,y)$. Figure 1(**b**) shows a representative hologram loaded onto SLM 1. The desired beam is spatially filtered and relayed to the output plane using a 4f imaging system consisting of a pair of Fourier lenses (L1-2). SLM 2 serves as the second conformal-mapping phase mask $\Phi_2(u,v)$ and is positioned at a distance $d = 60$ mm from the output plane of SLM 1. Figure 2(**c**) shows the hologram loaded onto SLM 2, which additionally incorporates a helical phase term. Finally, the far-field properties of the generated partially coherent beams are characterized via free-space propagation followed by focusing with a lens.

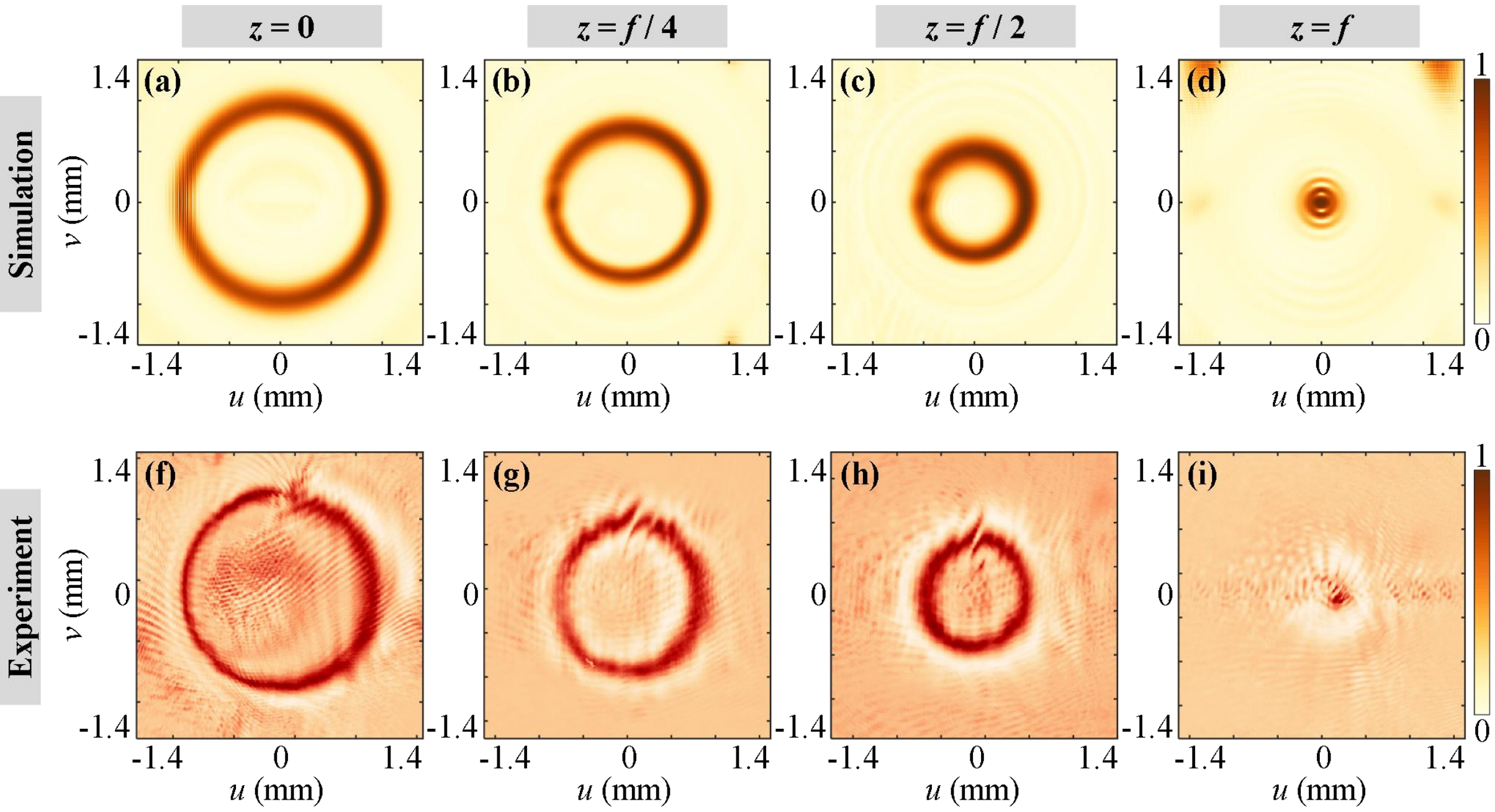

**Fig. 3.** (**a**)-(**d**) Simulated and (**f**)-(**i**) experimentally measured distributions of $|\mu(\vec{r}_1,\vec{r}_2)|^2$ for radially correlated partially coherent beams without a vortex structure at different propagation distances after focusing by a spherical lens. The parameters are set as $\delta_y = 0.1$ mm and $f = 400$ mm, the reference point is selected at the position corresponding to the maximum radial intensity. All results are obtained by averaging over 3000 realizations.

Figure 3 presents the simulated and experimentally measured distributions of $|\mu(\vec{r}_1,\vec{r}_2)|^2$ for radially correlated partially coherent beams without a vortex structure at different propagation distances after focusing by a spherical lens. At the source ($z = 0$), as shown in Figs. 3(**a**) and 3(**f**), the coherence structure of the radially correlated partially coherent beam exhibits a ring-shaped profile, wherein the experiment and simulation are in good agreement. As the propagation proceeds further ($z = f$), the radius of the coherent ring gradually decreases. Upon further propagation toward the focal region, the annular coherence structure progressively collapses and ultimately vanishes, resulting in the disappearance of the radial correlation distribution, as illustrated in Figs. 3(**b**)-(**d**) and 3(**g**)-(**i**).

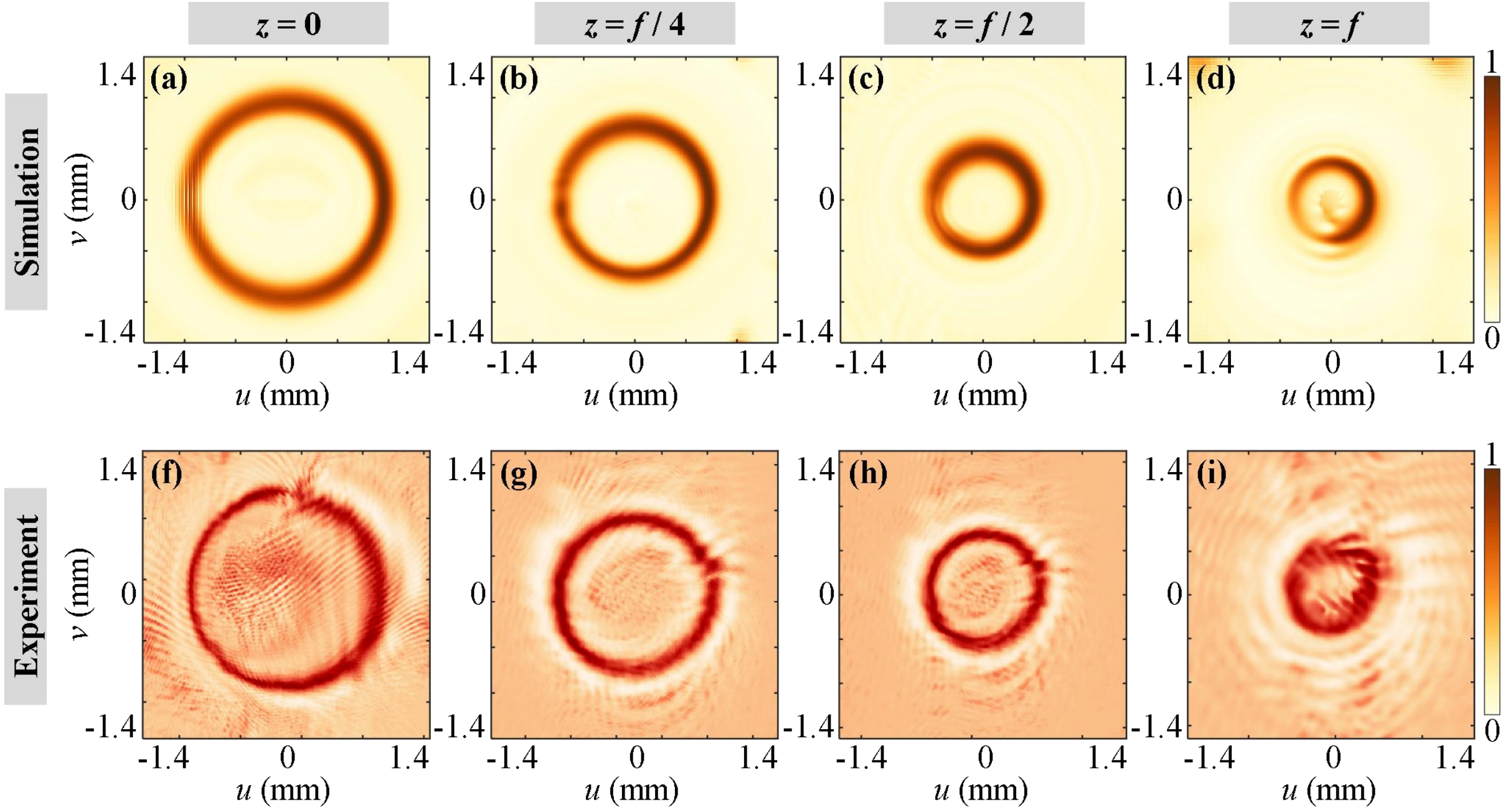

**Fig. 4.** (**a**)-(**d**) Simulated and (**f**)-(**i**) experimental measured distributions of $|\mu(\vec{r}_1,\vec{r}_2)|^2$ for radially correlated partially coherent beams with a vortex structure ( $\ell = 10$ ) at different propagation distances after focusing by a spherical lens. The parameters are set as $\delta_y = 0.1$ mm and $f = 400$ mm, the reference point is selected at the position corresponding to the maximum radial intensity. All results are obtained by averaging over 3000 realizations.

Unlike conventional GSM vortex beams, in which random wavefront fluctuations severely degrade the vortex structure and cause the orbital angular momentum (OAM) to diminish during propagation, the radially correlated partially coherent beam remains fully coherent along the azimuthal direction. Consequently, an additional helical phase can be effectively imposed on the beam, enabling the transfer and preservation of longitudinal OAM in the resulting field. In the experiment, a helical phase term was incorporated into the second phase mask displayed on SLM 2, thereby imprinting a helical wavefront onto the generated partially coherent beam, as illustrated in Fig. 2(**c**). Figure 4 shows the simulated and experimental results of $|\mu(\vec{r}_1,\vec{r}_2)|^2$ for radially correlated partially coherent beams with a deterministic vortex structure at different propagation distances after focusing by a spherical lens. It can be observed that, before reaching the focal region, the coherence structure exhibits a ring-shaped profile with a radius identical to those shown in Figs. 3(**b**)(**c**) and 3(**c**)(**h**). As the beam propagates toward the focus, however, the $|\mu(\vec{r}_1,\vec{r}_2)|^2$ continues to preserve its ring-shaped structure, in contrast to the collapse observed for Fig. 3(**d**)(**i**). This persistence arises from the deterministic vortex carried by the beam. To further characterize the vortex structure embedded in the radially correlated partially coherent beams, we experimentally measured the complex cross-spectral density using a Mach-Zehnder interferometer. Figure 5 displays the measured complex distributions of the cross-spectral density for various topological charges $\ell$. As can be seen, the cross-spectral density of radially correlated partially coherent beams carrying nonzero topological charges exhibits a ring-shaped distribution with a distinct dark core. Moreover, the radius of the dark core increases with the magnitude of the topological charge and the phase distribution exhibits $\ell$ azimuthal $2\pi$-phase twists, confirming that the beam carries a topological charge of $\ell$.

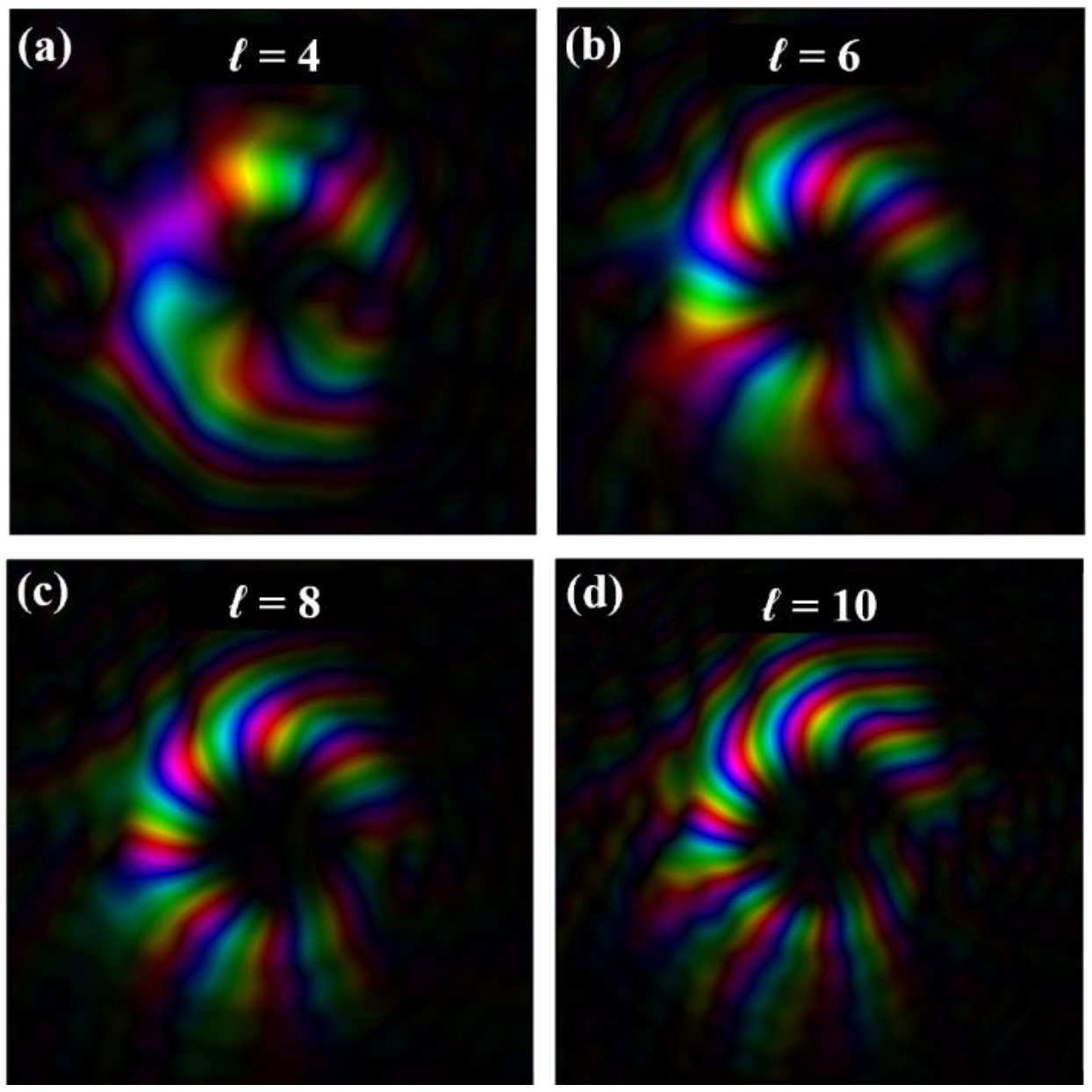

**Fig. 5.** Experimentally measured cross-spectral density functions of the generated radially correlated partially coherent beams with different vortex structures. The brightness and hue represent the magnitude and phase distributions, respectively. (**a**) $\ell = 4$; (**b**) $\ell = 6$; (**c**) $\ell = 8$ and (**d**) $\ell = 10$. Other parameters are identical to those used in Fig. 4.

**Conclusion**

In summary, we proposed and experimentally generated a series of radially correlated partially coherent beams with deterministic vortex structures by employing optical conformal mapping from Cartesian to log-polar coordinates. These beams exhibit a ring-shaped coherence distribution, characterized by partial coherence in the radial direction and high coherence along the azimuthal direction. Furthermore, we investigated their propagation dynamics both in the absence and presence of longitudinal orbital angular momentum. Numerical simulations and experimental measurements reveal that, without longitudinal orbital angular momentum, the ring-shaped coherence structure gradually collapses during propagation and eventually disappears in the far field. In contrast, the presence of longitudinal orbital angular momentum enables the beam to preserve its annular coherence profile over propagation. These findings provide new opportunities for statistical wavefront engineering and may facilitate the development of robust optical encoding, free-space information transmission, remote sensing, and ultrafast light-matter interaction applications.

**Funding.** National Natural Science Foundation of China (W2441005 [Y.C.], 12192254 [Y.C.], 12534014 [Y.C.], 12547149 [X.L.], 12374311 [C.L.]), National Key Research and Development Program of China (2022YFA1404800 [Y.C.]), Natural Science Foundation of Shandong Province (ZR2024QA216 [J.L.], ZR2025ZD21 [Y.C.], ZR2023YQ006 [C.L.]), Key Research and Development Program of Shandong Province (2024JMRH0105 [Y.C.]), Young Talent of Lifting engineering for Science and Technology in Shandong (SDAST2026 QTA028 [X.L.]), Taishan Scholar Project of Shandong Province (tsqn202312163 [C.L.]), China Postdoctoral Science Foundation [X.L.].

**Disclosures.** The authors declare no competing interests.

**Data availability.** The data that support the findings of this study are available from the corresponding authors upon reasonable request.

**Author contributions.** X.L. conceived the idea and directed the research; X-R.L., Q.C., K.Z., Y.Z. W.W. and Z.Z conducted the simulations and experiments; X-R.L., Y.Z. and X.L. completed the visualization. X-R.L., J.L., C.L., and X.L. performed the data analysis. J.L. and X.L. wrote the paper with input from X-R.L.; X.L., C.L., Y.C. and J.L. co-supervised the project. All authors contributed to the discussion and revision of the manuscript.